\documentclass[prb,preprint,aps]{revtex4-1} 

\usepackage{amsmath}  
\usepackage{amsfonts} 
\usepackage{graphicx} 
\usepackage{graphicx} 
\usepackage{caption,subcaption}
\usepackage{placeins}

\begin{document}


\title{Light Deviation around a Spherical Rotating Black Hole to Fifth Order. Lindstedt-Poincar\'e and Pad\'e Approximations.}


\author{Pablo Ruales}
\email{pmruales@estud.usfq.edu.ec}
\affiliation{Department of Physics, Universidad San Francisco de Quito, Diego de Robles y V\'ia Interoce\'anica }

\author{Carlos Mar\'in}
\email{cmarin@usfq.edu.ec} 
\affiliation{Department of Physics, Universidad San Francisco de Quito, Diego de Robles y V\'ia Interoce\'anica} 



\date{\today}
\begin{abstract}{Light deviation around a rotating black hole is calculated using the Kerr metric for both small and large deviation angles. For small angles the Lindstedt-Poincar\'e method is employed to get well-behaved solutions, as well as Pad\'e approximants. For large deviation angles numerical integration has been used.} 
\keywords{Light deviation \and Kerr \and spin \and Lindstedt \and Poincar\'e} 

\end{abstract}


\maketitle 
\section{Introduction}
\label{intro}

The formation of black holes is one of the most fascinating events in the universe, these objects are a manifestation of extreme space and time conditions where not even light can escape. `The General Theory of Relativity' (GTR) provides means to study such events, in particular, the solution to Einstein's field equations for a spherical rotating black hole was found by Roy Kerr in 1963. Nature has certain tendency to be represented by rotating bodies, in other words, most massive objects in the universe tend to acquire rotation. In the case of stellar-scale accumulations of mass, the space-time surrounding it is described by the Kerr metric. An object as massive as a black hole carrying spin can produce important space-time curvature along the axis of rotation, the geometry in the vicinity of a Kerr black hole is not easy to describe due to frame dragging effects which arise from the non-diagonal metric. In this paper we study the deviation of massless particles which enter the gravitational field of a rotating black hole. In the case of small deviation angles, the Lindstedt-Poincar\'e method provides a way to solve the equation of motion by developing a perturbative expansion. The solution is a series in a small parameter which we define as $\epsilon=\frac{r_c}{b}$, where $r_{c}$ is the critical radius and $b$ represents the impact parameter . The angle of deviation for photons was previously approximated with excellent precision for the Schwarzschild \cite{Rodriguez-Marin} and Reissner-Nordstr\"om \cite{Marin-Poveda} metrics using the Lindstedt-Poincar\'e and Pad\'e methods. Following a similar procedure as it was done in the aforementioned publications, we approximate the angle of deviation for photons that pass near a rotating black hole.

\section{The Kerr Metric}

\medskip{}

The Kerr metric describes space-time around a massive rotating body, without electric charge. If enough mass is accumulated, such as a black hole, the rotation produces notable effects in the geometry of space and time. The rotation around its own axis is given by the angular momentum of spin $ S_ {z} $ of the massive body M.The Kerr metric is given by  \cite{Misner,Ryder,Hobson,Chandrasekhar,tHooft,Ludvigsen,Marin}

\begin{eqnarray}
\label{mkerr1}
{d}{s}^{2} &=& c^{2}\left(dt\right)^{2} -\frac{\rho^{2}}{\Delta}\left(dr\right)^{2}-\rho^{2}\left(d\theta\right)^{2}-\left(r^{2}+a^{2}\right){\sin}^2{\theta}\left(d\phi\right)^{2}
\nonumber \\
&-& \frac{r_{s}r}{\rho^{2}}\left(c \ dt-a \ {\sin}^2{\theta} \ d\phi \right)^{2},
\end{eqnarray} 

\

\noindent where $a=\frac{S_{z}}{Mc}$, $\Delta=r^{2}-r_{s}r+a^{2}$ and $\rho^{2}=r^{2}+a^{2}{\cos}^{2}{\theta}$, with coordinates  $x^{0}=ct$, $x^{1}=r$, $ x^{2}=\theta$ and  $x^{3}=\phi$. $r_{s}=\frac{2GM}{c^{2}}$ is the Schwarzschild radius.

Throughout this paper, the motion of photons around the gravity source will be restricted to the equatorial plane, in other words, the trajectories considered will be in a plane perpendicular to the rotation axis of the source. The equations of motion in the Kerr metric impose great difficulty when attempting a solution, this is due to the non-diagonal nature of the space-time metric, thus, we set the polar coordinate $\theta=\frac{\pi}{2}$.With this value of $\theta$, the last equation can be written as

\begin{equation}\label{eq:mkerr}
\left(ds\right)^{2}=\left(c \text{ } dt\right)^2 \left(1-\frac{2 \mu}{r}\right)+\frac{4\mu a}{r}\text{ }c  \text{ }dt d\phi -\frac{r^2 }{\Delta }\left(dr\right)^2-\left(r^2+a^2+ \frac{2 \mu a^2}{r} \right) \left(d\phi\right) ^2,
\end{equation}

\noindent where $\mu=\frac{GM}{c^{2}}$.
Now, using the Lagrangian and Hamiltonian formalisms, we can find the equations of motion in the equatorial plane. First we define the Lagrangian:

$\mathcal{L}=\frac{1}{2}g_{\mu \nu }\dot{x}^{\mu }\dot{x}^{\nu }$ \text{  } and \text{  } $g^{\mu \nu }p_{\mu }p_{\nu }=\eta ^2$ \cite{},

\textup{where:}
$$\begin{array}{ll}
 \eta =m c & \text{for massive particles} \\
 \eta =0 & \text{for photons} \\
\end{array}$$

Hamilton's equations for massless particles are thus:

$$\begin{array}{ll}
p_t&=\frac{1}{c} \frac{\partial \mathcal{L}}{\partial \dot{t}}=\left(1-\frac{2\mu }{r}\right)c \dot{t}+\frac{2a \mu }{r}\dot{\phi }\\
p_{\phi }&=\frac{\partial \mathcal{L}}{\partial \dot{\phi }}=\frac{2a \mu }{r}c \dot{t}-\left(a^2+\frac{2a^2\mu }{r}+r^2\right)\dot{\phi }\left(a^2+\frac{2a^2\mu
}{r}+r^2\right)\\
p_r&=\frac{\partial \mathcal{L}}{\partial \dot{r}} =-\frac{r^2}{\Delta }\dot{r}\\
p_{\theta } &=0,
\end{array}$$

because the Lagrangian does not explicitly depend on the coordinates $\phi$ and $t$, then $p_t$ and $p_\phi$ are conserved along the corresponding geodesics, therefore:

$$\begin{array}{ll}
 p_t & \to \frac{E'}{c} \\
 p_{\phi } & \to -\mathit{h},
\end{array}$$

where $E'$ has units of energy per unit mass, and $\mathit{h}$ angular momentum per unit mass.

With these relations we are able to find the equations of motion which completely describe the orbits of photons in the equatorial plane:

\begin{equation}
c \dot{t}=c\frac{dt}{d\lambda }=\frac{1}{\Delta }\left[\left(a^2+\frac{2a^2\mu }{r}+r^2\right)\frac{E'}{c}-\frac{2a \mu }{r}\mathit{h}\right]
\end{equation}

\begin{equation}
\dot{\phi }=\frac{d\phi }{d\lambda }=\frac{1}{\Delta }\left[\left(\frac{2a \mu }{r}\right)\frac{E'}{c}+\left(1-\frac{2\mu }{r}\right)\mathit{h}\right]
\end{equation}

\begin{equation}
\left(\frac{dr}{d\lambda }\right)^2=\left(\frac{E'}{c}\right)^2+\frac{1}{r^2}\left(\left(a\frac{E'}{c}\right)^2-\mathit{h}^2\right)+\frac{1}{r^3}2\mu
\left(\mathit{h}-a\frac{E'}{c}\right)^2 ,
\end{equation}

\noindent where $\lambda$ is an affine parameter.

\subsection*{Equatorial trajectories of photons}

Having derived the differential equations that generally describe the motion of massless particles in the equatorial plane, to simplify the equations we introduce a parameter $b \equiv \frac{\mathit{h}c}{E'}$. This gives the following:

\begin{equation}\label{eq:radialeq}
\frac{1}{\mathit{h}^2}\left(\frac{dr}{d\lambda }\right)^2=\frac{1}{b^2}+\frac{1}{r^2}\left(\left(\frac{a}{b}\right)^2-1\right)+\frac{2\mu
}{r^3}\left(1-\frac{a}{b}\right)^2
\end{equation}

\begin{equation}\label{eq:angulareq}
\frac{d\phi }{d\lambda }=\frac{\mathit{h}}{\Delta }\left[\left(\frac{2 \mu }{r}\right)\frac{a}{b}+\left(1-\frac{2\mu }{r}\right)\right]
\end{equation}

\begin{equation}
\Longrightarrow \frac{d\phi }{dr}=\frac{\left[\left(\frac{2 \mu }{r}\right)\frac{a}{b}+\left(1-\frac{2\mu }{r}\right)\right]}{r^2\left(1-\frac{2\mu }{r}+\left(\frac{a}{r}\right)^2\right)\sqrt{\frac{1}{b^2}+\frac{1}{r^2}\left(\left(\frac{a}{b}\right)^2-1\right)+\frac{2\mu
}{r^3}\left(1-\frac{a}{b}\right)^2}}
\end{equation}

Considering the case where $r \longrightarrow \infty$, the equation is reduced to:

\[r^2\frac{d\phi }{dr}=-b,\]

\begin{figure}[h]
\begin{centering}
\includegraphics{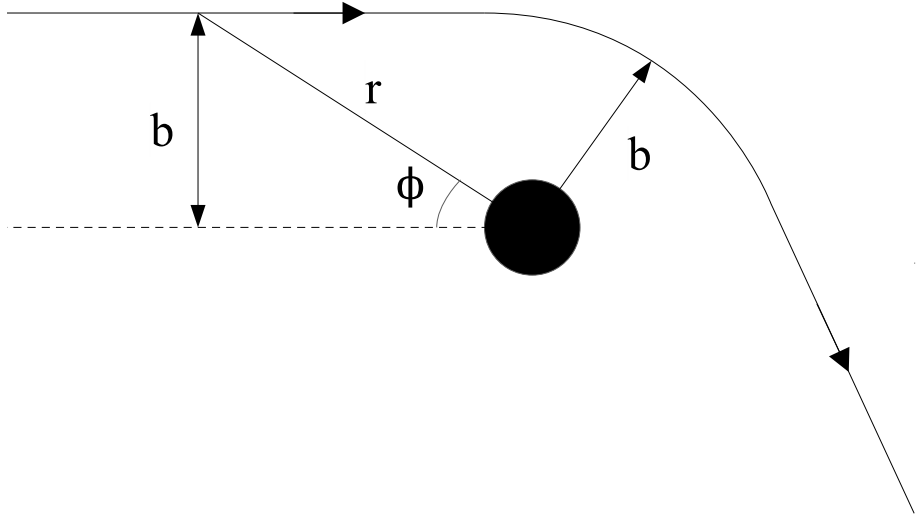}
\par\end{centering}
\caption{Impact parameter. ($b$ is the distance of maximum approach)\label{fig:impact}}
\end{figure}

From figure \ref{fig:impact}, we see that $\sin \phi=\frac{b}{r}$, which derivative is $d\phi  \cos\phi=-\frac{b}{r^2}dr$. For a small angle $\phi$ ($\cos \phi \approx 1$), we get the same differential equation as before ($r^2\frac{d\phi }{dr}=-b$).

Hence,  $b = \frac{\mathit{h}c}{E'}$ represents the impact parameter.

\section{Equation of the orbit}

From the equations of motion for null geodesics (equations \ref{eq:radialeq} and \ref{eq:angulareq}) we can derive the equation of the orbit by applying the transformation $u\left(\phi\right)=\frac{1}{r\left(\phi\right)}$. After lengthy algebra, we get the following equation:

\begin{equation}\label{eq:orbit1}
\frac{1}{u^4}\left(\frac{du}{d\phi }\right)^2\left[(2 \mu  u)\frac{a}{b}+1-2\mu  u\right]^2=\frac{\Delta
^2}{b^2}-\Delta ^2u^2\left(1-\frac{a^2}{b^2}-2\mu  u \left(1-\frac{a}{b}\right)^2\right)
\end{equation}

Now, as a first approximation we reduce equation \ref{eq:orbit1} to first order in spin by eliminating all higher order terms of $a=\mu s$. This is done to simplify further calculations once the perturbative and numeric approximations are applied, which become considerably complex when higher order terms are taken into account, and to sufficient precision in our results. 

We will encounter terms of the form $u^4\Delta ^2$, which can be approximated: $u^4\Delta ^2=\left(1-2\mu  u+a^2u^2\right)^2\approx  (1-2\mu  u)^2$, this reduces equation \ref{eq:orbit1} to:

\begin{equation}\label{eq:orbit2}
\left(\frac{du}{d\phi }\right)^2\left(1-2\mu  u+4\mu  u \frac{a}{b}\right)=(1-2\mu  u)\left(\frac{1}{b^2}-u^2\left(1-2\mu  u+4\mu  u \frac{a}{b}\right)\right),
\end{equation}
\noindent and defining
\begin{equation}
\sigma (u)\equiv 1-2\mu  u +4\mu  u \frac{a}{b}
\end{equation}

\begin{equation}
\Longrightarrow  \left(\frac{du}{d\phi }\right)^2\sigma (u)=(1-2\mu  u)\left(\frac{1}{b^2}-u^2\sigma (u)\right)
\end{equation}

Applying the derivative $\frac{d}{d\phi}$:

\begin{equation}
\frac{d^2u}{d\phi ^2}+\frac{(1-2\mu  u)}{2\sigma ^2}\left(\frac{1}{b^2}-u^2\sigma \right)\left(\frac{d\sigma }{du}\right)=-\frac{\mu }{\sigma }
\left(\frac{1}{b^2}-u^2\sigma \right)+\frac{(1-2\mu  u)}{2\sigma }\left(-2u \sigma -u^2\left(\frac{d\sigma }{du}\right)\right)
\end{equation}

\textup{where:} 

$$\begin{array}{cl}
 \frac{d\sigma}{du}&=-2\mu\left(1-2\frac{a}{b}\right)\\
 \sigma^{-1}(u)  & \approx 1+2\mu u\left(1-2\frac{a}{b}\right)+4\mu^2u^2\left(1-4\frac{a}{b}\right)\\
 \sigma^{-2}(u) & \approx 1+4\mu u\left(1-2\frac{a}{b}\right)-4\mu^2u^2\left(1-4\frac{a}{b}\right). \\
\end{array}$$

Finally, considering $a=\mu s$ to first order only, and $\mu u$ small (considering only terms up to $u^{6}$), after some lengthy algebra we arrive to the equation of the orbit:

\begin{equation}
\label{eq:orbitGeneral}
\begin{split}
\frac{d^2u}{d\phi ^2}+u=&-\frac{2\mu ^2s}{b^3}-\frac{8\mu ^3s}{b^3}u+3\mu  u^2-8\frac{\mu ^3}{b^2}u^2\left(2-9\frac{\mu  s}{b}\right)+8\frac{\mu ^4}{b^2}u^3\left(1-6\frac{\mu
 s}{b}\right)\\
 &+24\mu ^3u^4\left(1-6\frac{\mu  s}{b}\right)-16\mu ^4u^5\left(4-25\frac{\mu  s}{b}\right)+16\mu ^5u^6\left(1-8\frac{\mu  s}{b}\right)
\end{split}
\end{equation}

\section{Critical radius}

The critical radius of the orbit can be easily studied by analyzing the specific case of circular trajectories on equation \ref{eq:orbitGeneral}, taking only the first four terms of the right hand side of said equation:

\begin{equation*}
    \frac{d^2u}{d\phi ^2}+u = 3\mu  u^2\left(1-\frac{16}{3}\left(\frac{\mu}{b}\right)^2\right)-\frac{2\mu ^2s}{b^3}\left(1+4\mu u\right)
\end{equation*}

Defining $\delta \equiv \left(1-\frac{16}{3}\left(\frac{\mu}{b}\right)^2\right)$ and considering that $\frac{d^2u}{d\phi ^2}=0$ for circular orbits, we end up with a quadratic equation which can be easily solved:

\begin{equation*}
    u_c=\frac{\left(1+\frac{8\mu^3 s}{b^3}\right)+\left[\left(1+\frac{8\mu^3 s}{b^3}\right)^2+\frac{24\mu^3 s}{b^3}\delta\right]^{1/2}}{6\mu \delta},
\end{equation*}

and because $\frac{\mu}{b}$ is small:

\begin{equation*}
    \Longrightarrow u_c \approx \frac{1}{3\mu\delta},
\end{equation*}
then, the critical radius will be approximately $r_{c}=3 \mu \delta < 3\mu$ (because $\delta <1$).
Because the difference between $3 \mu \delta$ y $3\mu$ is small , we can use it as a small parameter $\epsilon \equiv \frac{r_c}{b}=\frac{3\mu}{b}$ (a non-dimensional small number) for our following perturbative expansions. This will be done on the next section.

Now let´s look at the stability of the circular orbits of photons in the Kerr metric. The stability condition is given by the nature of the second derivative of the effective potential, which we can derive from equation \ref{eq:radialeq}:

\begin{equation}\label{eq:Veffeq}
\frac{1}{\mathit{h}^2}\left(\frac{dr}{d\lambda }\right)^2 + V_{eff}'(r,b)=\frac{1}{b^2},
\end{equation}

where $V_{eff}'(r,b)\equiv \frac{1}{r^2}\left[1-\left(\frac{a}{b}\right)^2-\frac{2\mu}{r}\left(1-\frac{a}{b}\right)^2\right]$ is the effective potential for null geodesics.

For circular orbits, we have $\left(\frac{dr}{d\lambda }\right)=\left(\frac{d^2r}{d\lambda^2}\right)=0$, then, the equation of the orbit would take the form $V_{eff}'(r,b)=\frac{1}{b^2}$. Calculating the first derivative of the effective potential and setting said derivative equal to zero, we get that the critical radius is 

\begin{equation}
\label{eq:criticalrc}
r_{c}=3\mu\frac{1-\frac{a}{b}}{1+\frac{a}{b}},
\end{equation}
that implies $\left|\frac{a}{b}\right|<1$.

The second derivative of the effective potential with respect to $r$ is:

\begin{equation}\label{stability}
    \frac{d^2 V_{eff}'(r,b)}{dr^2}=\frac{6}{r^4}\left(1-\left(\frac{a}{b}\right)^2\right)\left(1-\frac{4\mu}{r}\frac{\left(1-\frac{a}{b}\right)}{\left(1+\frac{a}{b}\right)}\right).
\end{equation}

\noindent For example, taking $a= \mu s=0$ implies that  $r_{c}=3\mu$, and therefore $ \frac{d^2 V_{eff}'(r,b)}{dr^2}=-0.025 < 0$.

Introducing equation \ref{eq:criticalrc} in the expression on the effective potential we get the relation

\begin{equation}
\left(b+a\right)^{3}=27\mu^{2}\left(b-a\right),
\end{equation}
setting $y=b+a$, we can write $y^{2}-9\mu r_{c}=0$, from which $y=3\left(\mu r_{c}\right)^\frac{1}{2}$, and so $b=3\left(\mu r_{c}\right)^\frac{1}{2}-a$. Replacing the last expression in equation \ref{eq:criticalrc} we finally arrive to the equation for the critical radius (photon sphere radius) for a photon in the equatorial plane:
\begin{equation}
r_{c}^{3}-6\mu r_{c}^{2}+9\mu^{2}r_{c}-4\mu a^{2}=0,
\end{equation}
with solutions \cite{Hobson,Chandrasekhar}:

\begin{equation}
    r_{c\pm}=2\mu \left[1+ \cos \left(\frac{2}{3} \cos^{-1} \left(\pm \frac{a}{\mu}\right)\right)\right],
\end{equation}

where the `$+$' sign represents retrograde orbits and `$-$' is for direct orbits. For an extreme Kerr, for example, we obtain $r_{c+}=4\mu$ and $r_{c-}=\mu$, respectively.

\section{Perturbation theory}

\begin{figure}[ht]
\begin{centering}
\includegraphics{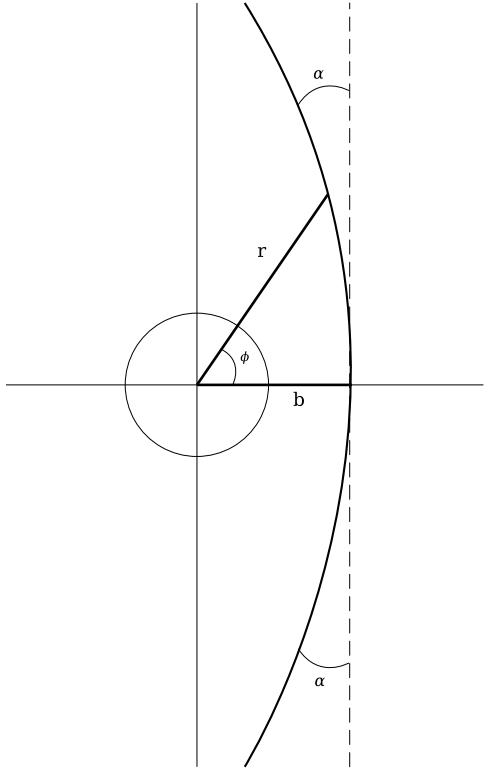}
\par\end{centering}
\caption{Angle of deviation $2\alpha$, where $b$ is the impact parameter.
\label{fig:deviation}}
\end{figure}

The equation for a photon orbiting a black hole in the Kerr metric is given by \ref{eq:orbitGeneral}, which has a polynomial nature. In a first attempt to solve this equation to find the angle of deviation for a photon that approaches from infinity (considering small deviations), we are going to try a perturbative treatment by Taylor series expansion. This can be done by expressing the equation of the orbit in terms of some small $\epsilon$, and a function that converges when the photon effectively escapes (returns to infinity, see figure \ref{fig:deviation}).

A converging function can be defined as follows:

$$\begin{array}{ll}
 V(\phi =0) & =\text{  }1 \\
 \frac{dV(\phi =0)}{d\phi } & =\text{  }0 \\
 |V(\phi )| & \leq \text{  }1 \\
\end{array}$$

where $V(\phi) \equiv \frac{b}{r(\phi)} = b u(\phi)$. Along with the small parameter $\epsilon \equiv \frac{r_c}{b} = \frac{3\mu}{b}$ the converging function $V(\phi)$ can be written as a power series in $\epsilon$:

\begin{equation}\label{eq:expansionV}
    V(\phi )=V_0(\phi )+\epsilon  V_1(\phi )+\epsilon ^2V_2(\phi )+\epsilon ^3V_3(\phi )+\epsilon ^4 V_4(\phi )+\ldots
\end{equation}

Then, the equation of the trajectory depicted in figure \ref{fig:deviation}, up to third order in $\epsilon$, would take the following form:

\begin{equation}\label{eq:trajectoryV}
\frac{d^2V}{d\phi ^2}+V=\epsilon  V^2-\frac{2}{9}\epsilon ^2s-\frac{8}{27}\epsilon ^3s V-\frac{16}{27}\epsilon ^3 V^2+\frac{8}{9}\epsilon ^3V^4
\end{equation}

This last equation sufficiently describes the behavior of photons that deviate due to the gravitational field caused by a rotating black hole. Notice that the spin parameter `$s$' appears in the second order term ($\epsilon^2$), thus, at least third order series expansion would be necessary to accurately approximate the angle of deviation once we solve this equation. As a first attempt to find a solution, we will expand \ref{eq:trajectoryV} as shown in equation \ref{eq:expansionV}:

\begin{equation}\label{eq:expandedV}
    \begin{split}
        & \left(\frac{d^2V_0(\phi )}{d\phi ^2} + \epsilon  \frac{d^2V_1(\phi )}{d\phi ^2}+\epsilon ^2\frac{d^2V_2(\phi )}{d\phi ^2}+\epsilon^3\frac{d^2V_3(\phi )}{d\phi ^2}+\ldots \right) +\\
        &+ \left(V_0(\phi )+\epsilon  V_1(\phi )+\epsilon ^2V_2(\phi )+\epsilon ^3V_3(\phi )+\ldots \right) = \\
        &\phantom{+} \epsilon \text{ }\left(V_0(\phi)+\epsilon  V_1(\phi )+\epsilon ^2V_2(\phi )+\epsilon ^3V_3(\phi )+\ldots \right){}^2 -\frac{2}{9}\epsilon^2s\\
        &-\frac{8}{27}\epsilon ^3s \left(V_0(\phi )+\epsilon  V_1(\phi )+\epsilon ^2V_2(\phi )+\epsilon ^3V_3(\phi )+\ldots \right) \\
        &-\frac{16}{27}\epsilon ^3 \left(V_0(\phi )+\epsilon  V_1(\phi)+\epsilon ^2V_2(\phi )+\epsilon ^3V_3(\phi )+\ldots \right){}^2 \\
        &+\frac{8}{9}\epsilon ^3\left(V_0(\phi )+\epsilon  V_1(\phi )+\epsilon ^2V_2(\phi)+\epsilon ^3V_3(\phi )+\ldots \right){}^4
    \end{split}
\end{equation}

Now that we have expanded the equation \ref{eq:trajectoryV} in a power series, it is possible to solve by separating \ref{eq:expandedV} by the order of $\epsilon$; this will lead to a system of equations that allows us to iteratively construct a solution for the function $V(\phi)$ of any desired order. Up to second order we have the following equations:

$$\begin{array}{ll}
 \text{For   } \epsilon^0: & \frac{d^2V_0}{d\phi ^2}+ V_0=0 \\
 \text{For   } \epsilon^1: & \frac{d^2V_1}{d\phi ^2}+ V_1=V_0{}^2 \\
 \text{For   } \epsilon^2: & \frac{d^2V_2}{d\phi ^2}+ V_2=-\frac{2}{9}s+2V_0V_1\\
\end{array}$$

The initial conditions are:

$$\begin{array}{lll}
 \text{for } V_0: & V_0(\phi =0)= 1 & \frac{dV_0(\phi =0)}{d\phi }= 0, \\
 \text{for } V_i: & V_i(\phi =0)= 0 & \frac{dV_i(\phi =0)}{d\phi }= 0, \\
 \text{where} & i \in \mathbb{N} \geq 1.\\
\end{array}$$

$$\begin{array}{ll}
    \Longrightarrow &  V_0(\phi )=\cos  (\phi )\\
     & V_1(\phi )=\frac{1}{6}(3-2\cos(\phi )-\cos(2\phi ))\\
     & V_2(\phi )=-\frac{1}{3}-\frac{2}{9}s+\left(\frac{29}{144}+\frac{2}{9}s\right)\cos(\phi)+\frac{1}{9}\cos(2\phi)+\frac{1}{48}\cos(3\phi)+\frac{5}{12}\phi  \sin(\phi )
\end{array}$$

These solutions build the function $V(\phi)$. Before attempting to find the angle of deviation, let's look at $V_2(\phi)$. The second order equation ($\epsilon^{2}$) contains a term which misbehaves in a series such as this one, a  term ($\frac{5}{12}\phi  \sin(\phi )$), it grows without bound with $\phi$, and occurs because the right-handed side of said equation contains terms proportional to the homogeneous solution of that equation: $a \, cos (\phi) + b \, \sin (\phi)$. When this happens, the solution contains terms that grow without bound, such as $\phi sin(\phi)$, called \textit{secular terms} \cite{Bush}. Thus, if we naively include that equation in $V(\phi)$, our solution is no longer bounded. Thus, we have to eliminate any and all secular term that arises to arrive at a well-behaved solution for $V(\phi)$. \\
One method to do this is due to Lindstedt and Poincar\'e as we shall see in the next section. Nonetheless, we shall calculate the angle of deviation, as depicted in figure \ref{fig:deviation}. First, to second order, the function $V(\phi )=V_0(\phi )+\epsilon  V_1(\phi )+\epsilon ^2V_2(\phi )$ can be put together as such:

\begin{equation}\label{eq:VsecondO}
    \begin{split}
        V(\phi)=& \cos \phi + \epsilon \left(\frac{1}{6}(3-2\cos(\phi )-\cos(2\phi ))\right)\\
        &+\epsilon^2 \left(-\frac{1}{3}-\frac{2}{9}s+\left(\frac{29}{144}+\frac{2}{9}s\right)\cos(\phi)+\frac{1}{9}\cos(2\phi)+\frac{1}{48}\cos(3\phi)+\frac{5}{12}\phi  \sin(\phi )\right),
    \end{split}
\end{equation}

remember that $V(\phi)=\frac{b}{r}$, the following condition must be met:

$$\text{as  } r \to \infty \Rightarrow V \to 0.$$

Therefore, when a photon is deviated, there must be an angle $\alpha$ that satisfies $V\left(\frac{\pi}{2}+\alpha\right)=0$. Replacing $\phi$ with $\frac{\pi}{2}+\alpha$ in equation \ref{eq:VsecondO} and solving for $\alpha$ gives the expression for the angle of deflection of light, once we eliminate all the higher order terms:

\begin{equation}
    \alpha =\frac{2}{3}\epsilon +\frac{\epsilon ^2}{9}\left(\frac{15\pi }{8}-2(s+1)\right)
\end{equation}

The total angle of deviation is $\Omega=2\alpha$:

\begin{equation}\label{eq:deviation2ndEpsilon}
    \Omega =\frac{4}{3}\epsilon +\frac{\epsilon ^2}{9}\left(\frac{15\pi }{4}-4(s+1)\right),
\end{equation}

given that $\epsilon=\frac{3\mu}{b}=\frac{3GM}{b c^2}$:

\begin{equation}\label{eq:deviation2nd}
    \Longrightarrow \Omega =\frac{4GM}{bc^2} +\left(\frac{GM}{bc^2}\right)^2\left(\frac{15\pi }{4}-4(s+1)\right)
\end{equation}

\section{Lindstedt-Poincar\'e}\label{LP}

We have successfully obtained the angle of deviation for a photon in the Kerr metric, to second order. This result is consistent with previous studies of second order corrections to the deflection angle for $s=0$. \cite{Rodriguez-Marin, Marin-Poveda}. The secular term that was mentioned previously does not affect the second order terms, it appears in third and higher orders. Therefore, to be able to calculate a third order solution we need to get rid of all secular terms that appear in the differential equations, to do this we employ the Lindstedt-Poincar\'e method \cite{Bush}. To eliminate the divergent terms from the higher order differential equations, an angle $\tilde{\phi}$ is defined as a power series in $\epsilon$:

\begin{equation}
    \tilde{\phi }=\phi  \left(1+\omega _1\epsilon +\omega _2\epsilon ^2+\omega _3\epsilon ^3+\ldots \right),
\end{equation}

where $\omega_i$ is a parameter that eliminates the secular term in the corresponding $i^{th}$ order equation. Rewriting \ref{eq:trajectoryV} in terms of $\tilde{\phi}$, to third order:

\begin{equation}\label{eq:taylor}
    \begin{split}
        \left(1+\omega _1\epsilon +\omega _2\epsilon ^2+\omega _3\epsilon ^3\right){}^2\frac{d^2V\left(\tilde{\phi}\right)}{d\tilde{\phi}^2}+V\left(\tilde{\phi }\right)=& \epsilon  V^2\left(\tilde{\phi}\right)-\frac{2}{9}\epsilon ^2s-\frac{8}{27}\epsilon ^3s\text{ } V\left(\tilde{\phi }\right) \\
        &-\frac{16}{27}\epsilon^3 V^2\left(\tilde{\phi }\right)+\frac{8}{9}\epsilon^3 V^4\left(\tilde{\phi }\right)
    \end{split}
\end{equation}

Performing the expansion $V(\tilde{\phi},\epsilon)=V_0(\tilde{\phi} )+\epsilon  V_1(\tilde{\phi} )+\epsilon ^2V_2(\tilde{\phi} )+\epsilon ^3V_3(\tilde{\phi} )+\ldots$ in the equation above, produces the following system of equations:

$$\begin{array}{lll}
 \text{For   } \epsilon^0: & \frac{d^2V_0}{d\phi ^2}+ V_0&=0 \\
 \text{For   } \epsilon^1: & \frac{d^2V_1}{d\tilde{\phi }^2}+ V_1&=V_0{}^2-2\omega _1V_0\text{''} \\
 \text{For   } \epsilon^2: & \frac{d^2V_2}{d\tilde{\phi }^2}+ V_2&=-\frac{2}{9}s+2V_0V_1-\left(\omega _1{}^2+2\omega _2\right)V_0\text{''}-2\omega _1V_1\text{''} \\
 \text{For   } \epsilon^3: & \frac{d^2V_3}{d\tilde{\phi }^2}+ V_3&=-\frac{8s}{27}V_0-\frac{16}{27} V_0{}^2+\frac{8}{9} V_0{}^4+V_1{}^2+2 V_0 V_2-\left(2 \omega _1\omega _2 +2 \omega _3\right) V_0''\\
 &&\phantom{=}-\left(\omega _1^2+2 \omega _2\right) V_1''-2 \omega _1 V_2''
\end{array}$$

To solve the equations above, the same initial conditions have to be considered:

$$\begin{array}{lll}
 \text{for } V_0: & V_0(\phi =0)= 1 & \frac{dV_0(\phi =0)}{d\phi }= 0, \\
 \text{for } V_i: & V_i(\phi =0)= 0 & \frac{dV_i(\phi =0)}{d\phi }= 0, \\
 \text{where} & i \in \mathbb{N} \geq 1,\\
\end{array}$$

\noindent before arriving at the solutions, the values of $\omega_i$ have to be determined so the secular terms are eliminated.

$$\begin{array}{l}
     \omega_1 = 0\\
     \omega_2 = -\frac{5}{12}\\
     \omega_3 = \frac{1}{54}(15+20s)\\
\end{array}$$

Introducing these parameters into the solutions to the differential equations, we obtain well behaved solutions with the divergent terms removed:

$$\begin{array}{ll}
     V_0\left(\tilde{\phi }\right)&=\cos  (\phi )\\
     V_1\left(\tilde{\phi }\right)&=\frac{1}{6}(3-2\cos(\phi )-\cos(2\phi ))\\
     V_2\left(\tilde{\phi }\right)&=\frac{1}{144} \left(-48-32 s+29 \cos(\phi)+32 s \cos(\phi)+16
\cos(2\phi)+3 \cos(3\phi)\right) \\
    V_3\left(\tilde{\phi }\right)&=\frac{1}{6480}\left(3615+1440 s-1657 \cos(\phi)-960 s \cos(\phi)-1760 \cos(2\phi)-480 s \cos(2\phi)\right.\\
    &\left.\phantom{=} -135 \cos(3\phi)-63 \cos(4\phi)\right)
\end{array}$$

The solution is $V(\tilde{\phi} ,\epsilon)=V_0(\tilde{\phi} )+\epsilon  V_1(\tilde{\phi} )+\epsilon ^2V_2(\tilde{\phi} )+\epsilon ^3V_3(\tilde{\phi})$, to third order:

\begin{equation}
    \begin{split}
        V(\tilde{\phi},\epsilon ) &= \cos  (\tilde{\phi} ) + \frac{\epsilon}{6}(3-2\cos(\tilde{\phi} )-\cos(2\tilde{\phi} )) + \frac{\epsilon^2}{144} \left(-48-32 s+29 \cos(\tilde{\phi})+32 s \cos(\tilde{\phi})\right.\\
        &+ \left. 16\cos(2\tilde{\phi})+3 \cos(3\tilde{\phi})\right)+\frac{\epsilon^3}{6480}\left(3615+1440 s-1657 \cos(\tilde{\phi})-960 s \cos(\tilde{\phi})\right.\\
        &\left. -1760 \cos(2\tilde{\phi})-480 s \cos(2\tilde{\phi})-135 \cos(3\tilde{\phi})-63 \cos(4\tilde{\phi})\right)
    \end{split}
\end{equation}

In an attempt to simplify the previous equation, we rewrite it in terms of powers of cosine, this allows for easier replacement of values of $\tilde{\phi}$:

\begin{equation}\label{eq:lindstedtSol}
    \begin{split}
        V(\tilde{\phi},\epsilon ) &= \cos \tilde{\phi}+\frac{1}{3} \left(2-\cos( \tilde{\phi})-\cos^2( \tilde{\phi})\right) \epsilon +\frac{1}{36} \left(-16-8 s+5 \cos( \tilde{\phi})+8 s \cos( \tilde{\phi})\right. \\
        &\left. +8 \cos^2( \tilde{\phi})+3 \cos^3( \tilde{\phi})\right) \epsilon ^2+\left(\frac{332}{405}+\frac{8 s}{27}-\frac{313 }{1620}\cos( \tilde{\phi})-\frac{4s}{27} \cos( \tilde{\phi})\right.\\
        &\left. -\frac{377 }{810}\cos^2( \tilde{\phi})-\frac{4s}{27}\cos^2( \tilde{\phi})-\frac{1}{12}\cos^3( \tilde{\phi})-\frac{7}{90}\cos^4( \tilde{\phi})\right) \epsilon ^3
    \end{split}
\end{equation}

The solution to the equation of motion provides means to find the angle of deviation of a deflected photon, remember that the condition $V(\frac{\pi}{2}+\tilde{\alpha})=0$ must be met. As it can be observed in equation \ref{eq:lindstedtSol}, we would need to solve an equation of polynomial nature with a sine function of increasing degree; evidently, this becomes very troublesome to deal with at higher orders. Therefore, the $\sin(\alpha)$ function can be expressed in terms of a power series with $\epsilon$ as a leading term (this allows the sine function to behave properly at small angles).

\begin{equation}
    \Longrightarrow \sin \left(\tilde{\alpha }\right)=\epsilon  \chi _1+\epsilon ^2\chi _2+\epsilon ^3\chi _3+\ldots
\end{equation}

First, replacing $\tilde{\phi}=\frac{\pi}{2}+\tilde{\alpha}$, we get the following equation:

\begin{equation}
    \begin{split}
        0 &= -\sin \tilde{\alpha}+\frac{\epsilon}{3} \left(2+\sin \tilde{\alpha}-\sin^2 \tilde{\alpha}\right)+\frac{\epsilon ^2}{36}  \left(-16-8s-5 \sin \tilde{\alpha}-8 s \sin \tilde{\alpha} \right.\\
        &\left. +8 \sin^2 \tilde{\alpha}-3 \sin^3 \tilde{\alpha}\right)+\epsilon ^3 \left(\frac{332}{405}+\frac{8s}{27}+\frac{313 }{1620}\sin \tilde{\alpha}+\frac{4s}{27} \sin \tilde{\alpha}-\frac{377 }{810}\sin^2 \tilde{\alpha} \right.\\
        &\left. -\frac{4s}{27} \sin^2 \tilde{\alpha}+\frac{1}{12}\sin^3 \tilde{\alpha}-\frac{7}{90}\sin^4 \tilde{\alpha}\right)
    \end{split}
\end{equation}

Now, applying the series expansion of the sine function, we can find the $\chi_i$ coefficients, which construct the angle of deviation.

\begin{equation}
    \begin{split}
        0 &= \left(\frac{2}{3}-\chi _1\right) \epsilon +\frac{1}{9} \left(-4-2 s+3 \chi _1-9 \chi _2\right) \epsilon ^2+\frac{1}{1620}\left(1328+480 s-225\chi _1\right.\\
        &\left. \phantom{=}-360 s \chi _1-540 \chi _1^2+540 \chi _2-1620 \chi _3\right) \epsilon ^3
    \end{split}
\end{equation}

$$\begin{array}{ll}
    \Longrightarrow &  \chi _1\to \frac{2}{3}\\
     & \chi _2\to -\frac{2}{9} (1+s)\\
     & \chi _3\to \frac{1}{810} (409+60 s)
\end{array}$$

\begin{equation}
    \Longrightarrow \sin \tilde{\alpha} = \frac{2 \epsilon }{3}-\frac{2}{9} (1+s) \epsilon ^2+\frac{1}{810} (409+60 s) \epsilon ^3 + \ldots
\end{equation}

To find $\tilde{\alpha}$, we need to apply the inverse sine function, this of course has to be expanded in its Taylor series to the desired order. Since we are working up to third order, the expansion is as follows:

\begin{equation*}
    \arcsin x = x +\frac{x ^3}{6}+O[x ]^5
\end{equation*}

\begin{equation}
    \Longrightarrow \tilde{\alpha}=\frac{2 \epsilon }{3}-\frac{2}{9} (1+s) \epsilon ^2+\left(\frac{449}{810}+\frac{2 s}{27}\right) \epsilon ^3
\end{equation}

Remember that the Lindstedt-Poincar{\' e} method expands the angle as a power series, thus, to find the actual deviation angle we need to revert said transformation.

\begin{equation*}
    \tilde{\phi }=\phi  \left(1+\omega _1\epsilon +\omega _2\epsilon ^2+\omega _3\epsilon ^3 \right)
\end{equation*}

\begin{equation*}
    \Longrightarrow \frac{\pi}{2}+\tilde{\alpha} = \left(\frac{\pi}{2}+\alpha\right) \left(1-\frac{5}{12}\epsilon^2+\frac{1}{54}(15+20s)\epsilon^3\right)
\end{equation*}

\begin{equation*}
    \frac{\pi}{2}+\left(\frac{2 \epsilon }{3}-\frac{2}{9} (1+s) \epsilon ^2+\left(\frac{449}{810}+\frac{2 s}{27}\right) \epsilon ^3\right) = \left(\frac{\pi}{2}+\alpha\right) \left(1-\frac{5}{12}\epsilon^2+\frac{1}{54}(15+20s)\epsilon^3\right)
\end{equation*}

\begin{equation}
    \Longrightarrow \alpha = \frac{2 \epsilon }{3}+\frac{1}{72} (-16+15 \pi -16 s) \epsilon ^2+\frac{(1348-225 \pi +120 s-300 \pi  s) \epsilon ^3}{1620}
\end{equation}

Finally, we can find the total deviation:

\begin{equation*}
    \Omega=2\alpha=\frac{4 \epsilon }{3}+\frac{\epsilon ^2}{36} (-16+15 \pi -16 s) +\frac{\epsilon ^3}{810} (1348-225 \pi +120 s-300 \pi  s) 
\end{equation*}

\begin{equation}\label{eq:omega}
    \Omega = \frac{4 \epsilon }{3}+\frac{\epsilon ^2}{9} \left(\frac{15 \pi}{4} -4(1+ s)\right) +\frac{\epsilon ^3}{27} \left(\frac{674}{15}-\frac{15\pi}{2} +(4 -10 \pi)  s\right)
\end{equation}

Observe that the first two terms in \ref{eq:omega} are in agreement with the ones given by equation \ref{eq:deviation2nd}.

\section{Results}

\begin{figure}[htb]
    \centering
\begin{subfigure}{0.5\textwidth}
  \includegraphics[width=\linewidth]{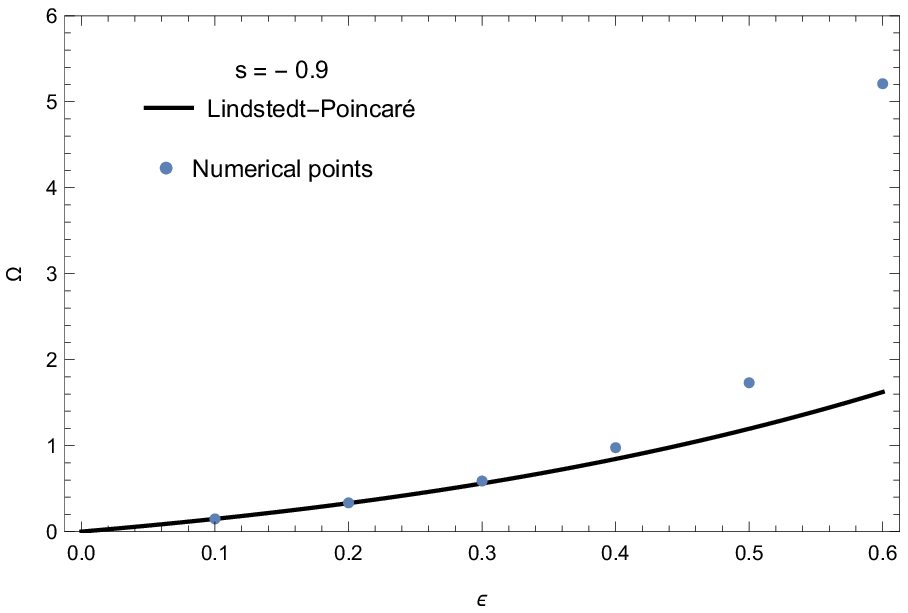}
  \caption{}
  \label{fig:-0.9}
\end{subfigure}\hfil
\begin{subfigure}{0.5\textwidth}
  \includegraphics[width=\linewidth]{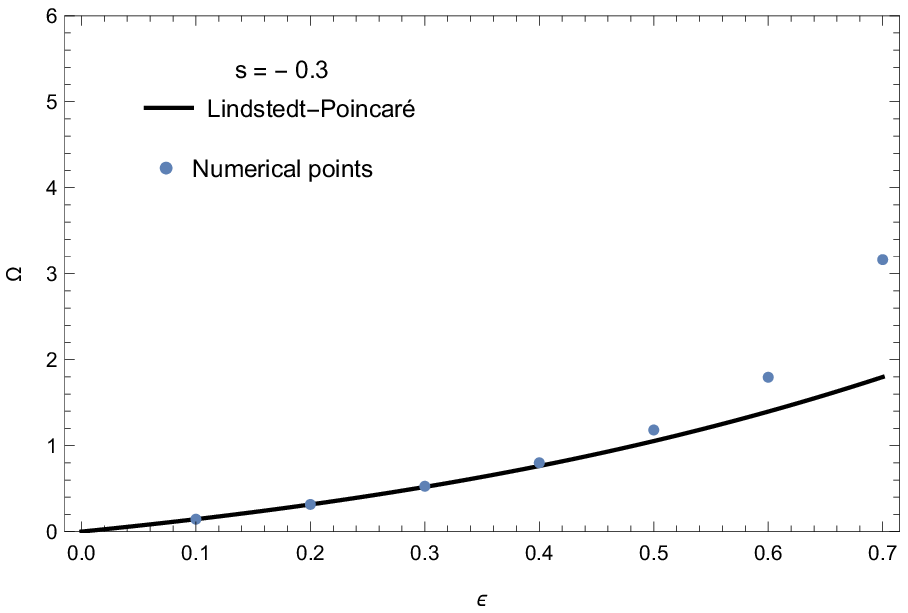}
  \caption{}
  \label{fig:-0.3}
\end{subfigure}\hfil

\medskip
\begin{subfigure}{0.5\textwidth}
  \includegraphics[width=\linewidth]{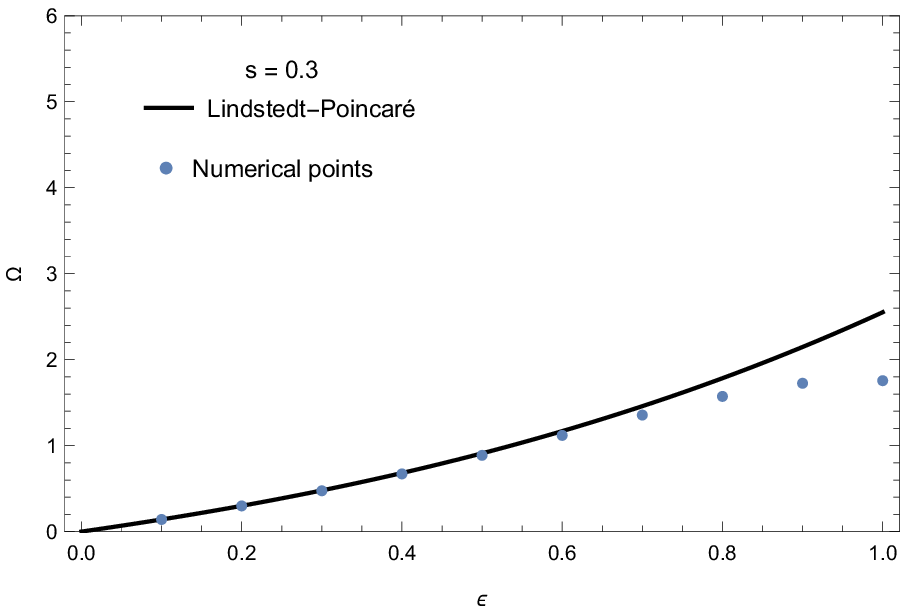}
  \caption{}
  \label{fig:0.3}
\end{subfigure}\hfil
\begin{subfigure}{0.5\textwidth}
  \includegraphics[width=\linewidth]{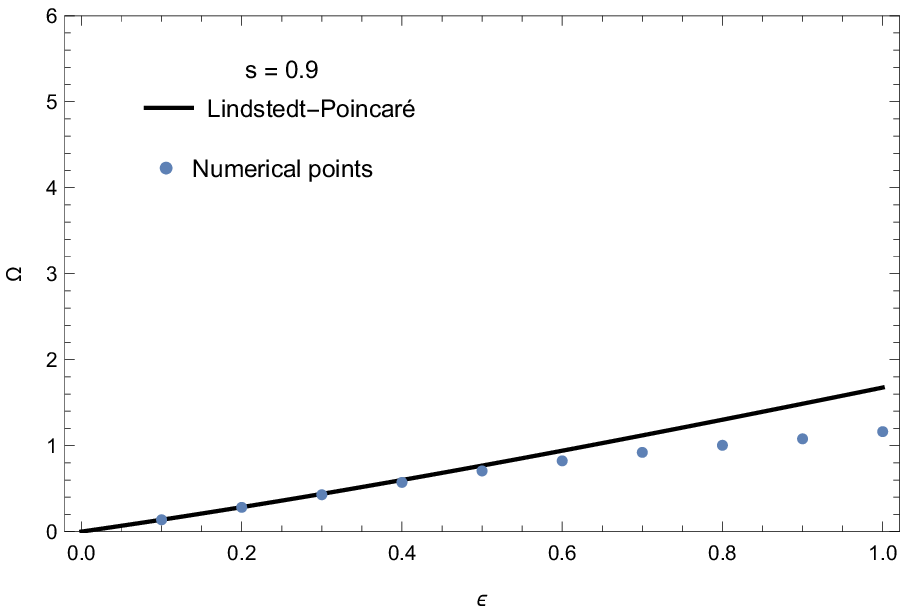}
  \caption{}
  \label{fig:0.9}
\end{subfigure}\hfil
\caption{Angle of deviation as a function of $\epsilon$ for different spin parameters. The solid line represents the solution given by equation \ref{eq:omega}, and the points are the solutions to equation \ref{eq:orbit2}. These plots show the Lindstedt-Poincar\'e and numerical solutions for spin parameters: 
(a) $s=-0.9$, (b) $s=-0.3$, (c) $s=0.3$ and (d) $s=0.9$. }
\label{fig:lindst}
\end{figure}

Now we will compare the results obtained in the previous section to the numerical solution to equation \ref{eq:orbit2}. Plotting equation \ref{eq:omega} for different values of spin and $\epsilon$, allows us to observe how the Lindstedt-Poincar\'e method approximates to the deviation angle given by the values that come from solving equation \ref{eq:orbit2} numerically. Figure \ref{fig:lindst} shows the behavior of a photon's deviation angle as it approaches a rotating black hole at different distances from its center, note that the deviation angle strongly depends on the spin parameter that the black hole holds. The solutions from the Lindstedt-Poincar\'e method, represented in figure \ref{fig:lindst}, are compared to the numeric solution to equation \ref{eq:orbit2}. Note that the error of the perturbative method increases as the particle approaches the black hole, specially for figures (\ref{fig:0.3}) and (\ref{fig:0.9}). This proves that higher orders of the Lindstedt-Poincar\'e solutions are necessary to accurately describe the behavior of light's deviation near a rotating black hole. In the next section, the Pad\'e approximants method will be employed in an attempt to get a better approximation for the angle of deviation. Finally, on all figures the $y$ axis represents the deviation angle, and the $x$ axis $\epsilon$ (which tells us how close to the black hole the particle passes). 

\newpage

\section{Pad\'e approximants}\label{pade}

The method of Pad\'e \cite{Cuyt} will be employed to find a rational approximation of the deviation angle, which was calculated as a power series. This method has been used to study the light deviation near Schwarzschild and Reissner-Nordstrom black holes \cite{Rodriguez-Marin, Marin-Poveda}, and also in Cosmology \cite{Luongo,Liu}. The Pad\'e approximant is defined as follows:

Given a power series:

\begin{equation*}
    f(x)=\sum _{k=0}^{\infty } c_kx^k
\end{equation*}

The rational function of order $[m/n]$:

\begin{equation*}
    R^{[m/n]}(x)=\frac{a_0+a_1x+\ldots +a_mx^m}{1+b_1x+\ldots +b_nx^n},
\end{equation*}

must match the power series $f(x)$ up to its derivative of order $m+n$:

\[R(0)=f(0),R'(0)=f'(0),\ldots ,
R^{(m+n)}(0)=f^{(m+n)}(0).\]

To find the coefficients of the polynomials in $R^{[m/n]}$, the following system of equations is used \cite{Cuyt}, which satisfies the conditions stated above.

$$\begin{array}{l}
     c_0b_0=a_{0},\\
     c_1b_0+c_0b_1=a_1,\\
     \ldots\\
     c_mb_0+c_{m-1}b_1+\ldots +c_{m-n}b_n=a_{m},\\
     c_{m+1}b_0+c_mb_1+\ldots +c_{m-n+1}b_n=0,\\
     \ldots \\
     c_{m+n}b_0+c_{m+n-1}b_1+\ldots +c_mb_n=0.\\
     \left(\text{If  } i<0 \Rightarrow c_i=0\right)
\end{array}$$

For the deviation angle calculated with the Lindstedt-Poincar\'e method (\ref{eq:omega}), applying the procedure above, we obtain the following Pad\'e approximants:

$$\begin{array}{l}
     \Omega^{[1/1]}(\epsilon,s)\equiv R^{[1/1]}(\epsilon,s)=\frac{64 \epsilon }{48-15 \pi  \epsilon +16 (1+s) \epsilon }\\
     \Omega^{[1/2]}(\epsilon,s)\equiv R^{[1/2]}(\epsilon,s)=\frac{46080 \epsilon }{34560+\epsilon  \left(3375 \pi ^2 \epsilon
+1200 \pi  (-9+2 s \epsilon )+128 (90-307 \epsilon +30 s (3+\epsilon +s \epsilon ))\right)}\\
     \Omega^{[2/1]}(\epsilon,s)\equiv R^{[2/1]}(\epsilon,s)=\frac{\epsilon  \left(3375 \pi ^2 \epsilon +1200 \pi(9+2 s \epsilon )+128 (-90-307 \epsilon +30 s (-3+\epsilon +s \epsilon ))\right)}{900 \pi  (9+(6+8 s) \epsilon )-96 (90+337 \epsilon +30 s (3+\epsilon))}
\end{array}$$

To compute higher order Pad\'e approximants, it is necessary to calculate the angle of deviation with the Lindstedt-Poincar\'e method of order $n+m$. To accurately approximate the angle of deviation, it was found that at least a fifth order solution is necessary.

\section{Analysis and Numerical Tests}
\label{analysis}

The Pad\'e approximants calculated numerically can be compared to the numerical solution of the equation, and to the results from the Lindstedt-Poincar\'e method. Clearly, the Pad\'e approximants provide more accurate results, especially as $\epsilon$ increases, compared to the Lindstedt-Poincar\'e solution. This means that the rational approximation given by Pad\'e provides means to find a better fit for the solution to equation (\ref{eq:orbit2}) than equations (\ref{eq:lindstedtSol}) and (\ref{eq:lindstedt5}). Now, determining which Pad\'e approximant is best for each case is a manual process that involves calculating the statistical error between each of the numerical points and the corresponding Pad\'e values, then taking the mean error. This allows us to choose the Pad\'e approximant that fits the numerical points best, overall.

The Pad\'e approximants of fifth order used in figure \ref{fig:lindst-pade} are too long to be shown here, they are presented in Appendix A. The expression for the angle of deviation calculated with the Lindstedt-Poincar\'e method to fifth order, following the same procedure as in section \ref{LP} is as follows:

\begin{equation}\label{eq:lindstedt5}
    \begin{split}
        \Omega_{5} (\epsilon,s)&=\frac{4 \epsilon }{3}+\frac{1}{36} (-16+15 \pi -16 s) \epsilon ^2+\frac{1}{810} (1348-225 \pi +120 s-300 \pi  s) \epsilon ^3\\
    &+\frac{\left(-176000+44235\pi -191616 s+14400 \pi  s+7680 s^2\right) \epsilon ^4}{77760}\\
    &+\frac{\left(1489396-427245 \pi +1564192 s-484680 \pi  s+161280 s^2\right) \epsilon^5}{408240}
    \end{split}
\end{equation}

\newpage

\begin{figure}[ht]
\begin{centering}
\includegraphics[width=0.6\linewidth]{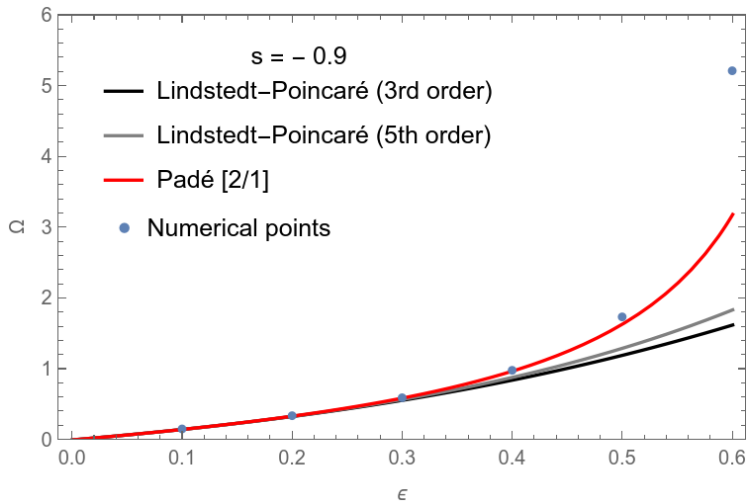}
\par\end{centering}
\caption{Angle of deviation as a function of $\epsilon$ for spin parameter: $s=-0.9$. The solid lines are given by equations (\ref{eq:lindstedtSol}), (\ref{eq:lindstedt5}) and $\Omega^{[2/1]}=R^{[2/1]}$, compared to the numerical solution to equation (\ref{eq:orbit2}). The statistical error for the Pad\'e approximation is $e=7.60\%$.}
\label{fig:-0.9LPP}
\end{figure}

\begin{figure}[htb]
    \centering 
\begin{subfigure}{0.45\textwidth}
  \includegraphics[width=\linewidth]{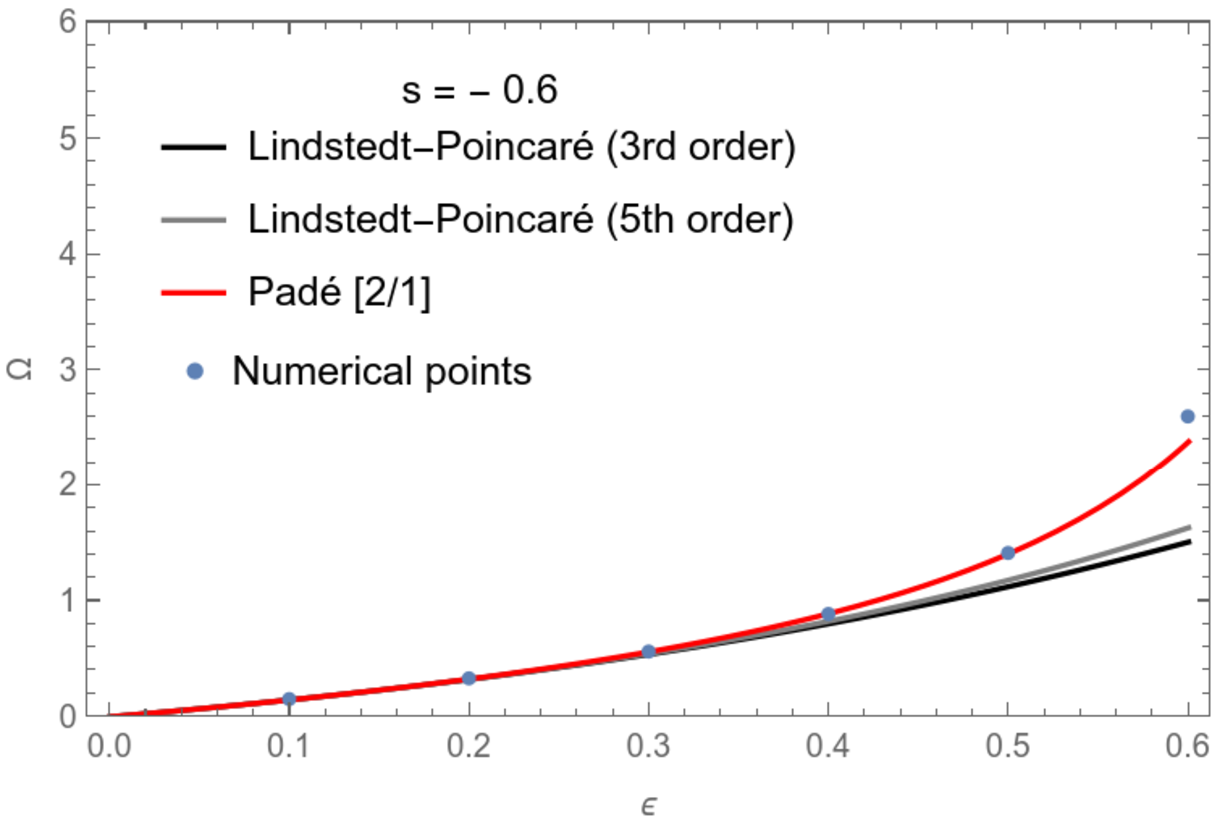}
  \caption{}
  \label{fig:-0.6LPP}
\end{subfigure}\hfil 
\begin{subfigure}{0.45\textwidth}
  \includegraphics[width=\linewidth]{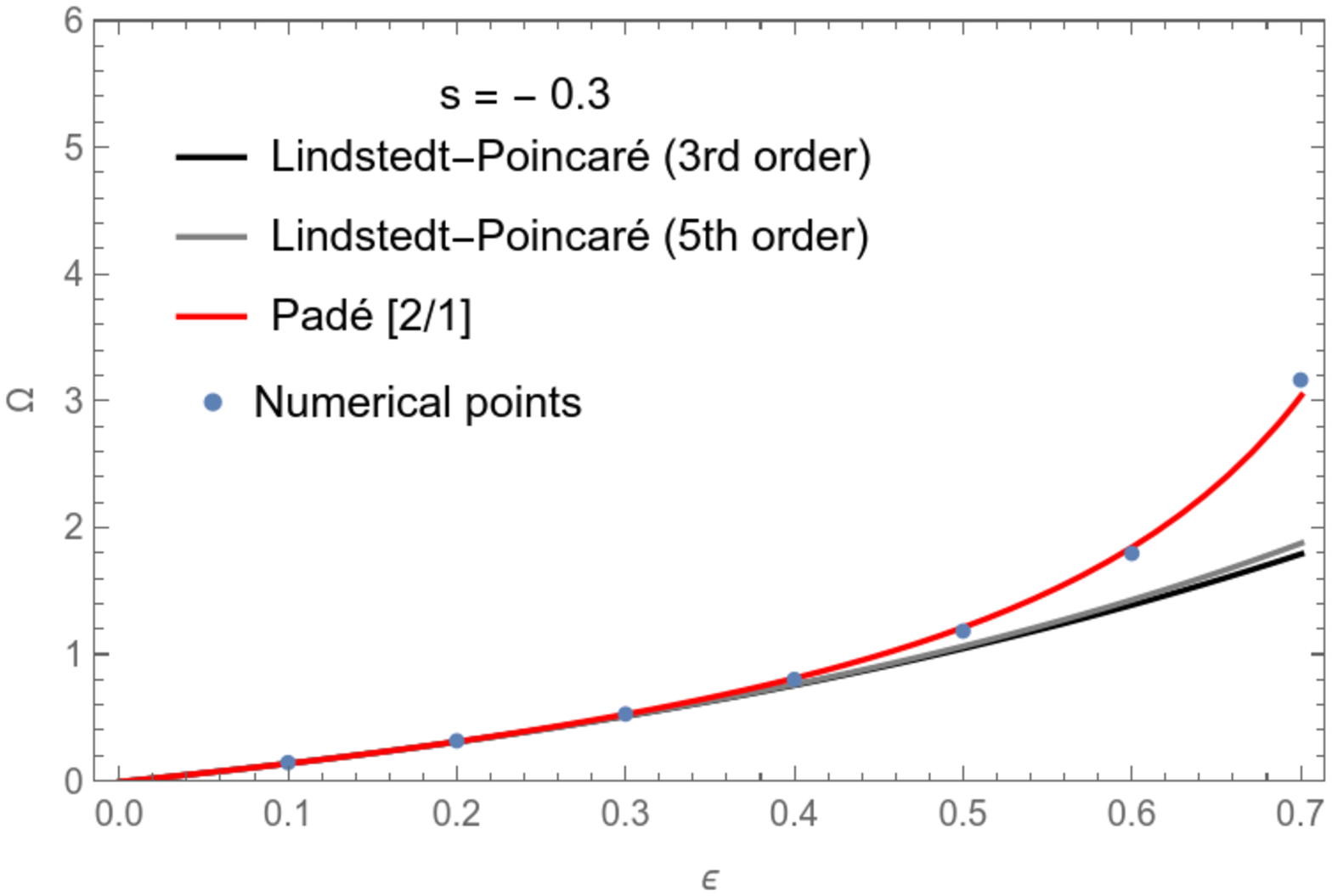}
  \caption{}
  \label{fig:-0.3LPP}
\end{subfigure}\hfil 

\medskip
\begin{subfigure}{0.45\textwidth}
  \includegraphics[width=\linewidth]{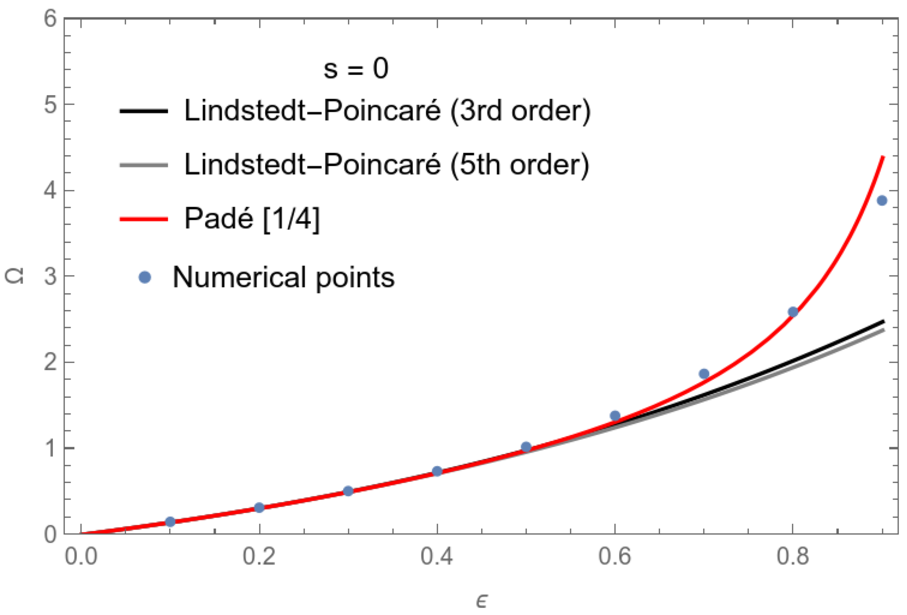}
  \caption{}
  \label{fig:0LPP}
\end{subfigure}\hfil 
\begin{subfigure}{0.45\textwidth}
  \includegraphics[width=\linewidth]{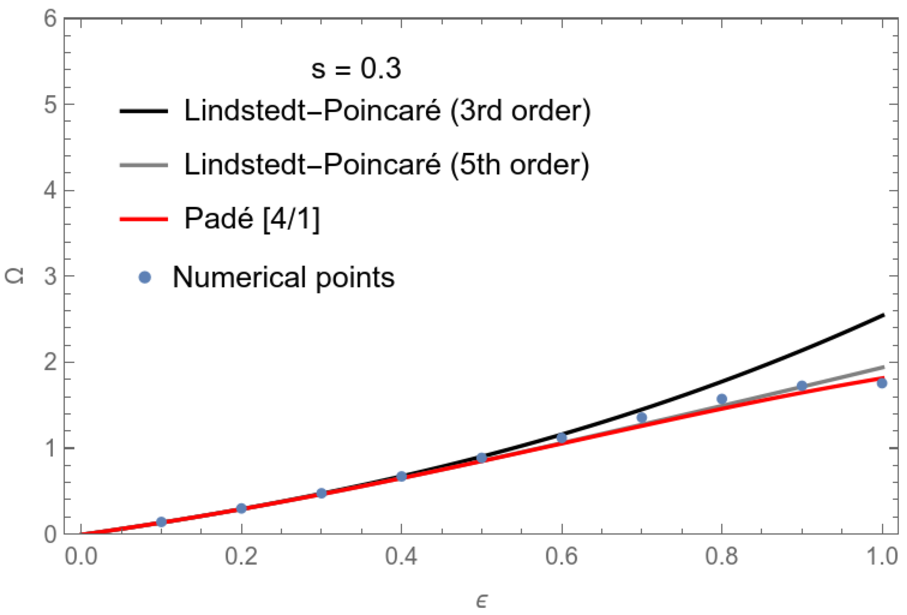}
  \caption{}
  \label{fig:0.3LPP}
\end{subfigure}\hfil 
\medskip
\begin{subfigure}{0.45\textwidth}
  \includegraphics[width=\linewidth]{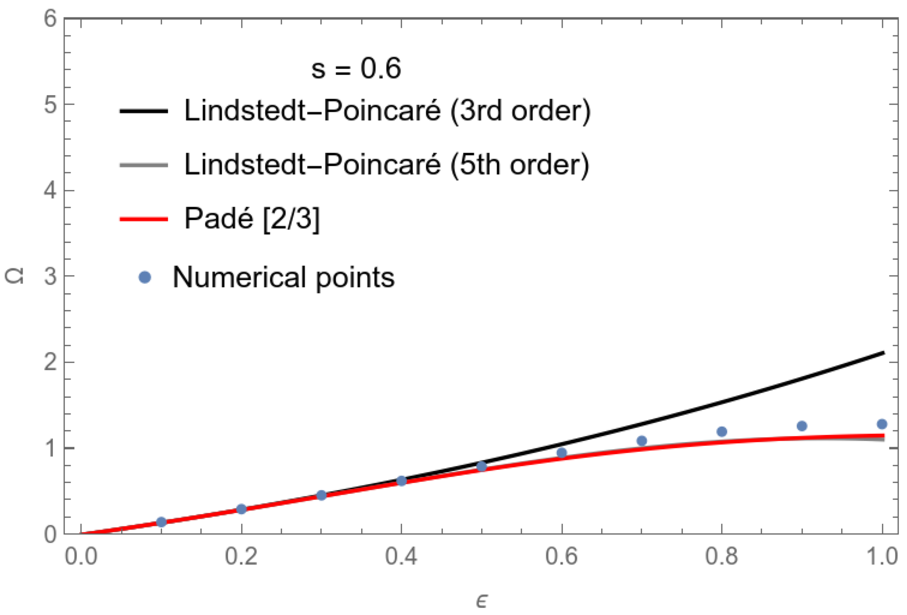}
  \caption{}
  \label{fig:0.6LPP}
\end{subfigure}\hfil 
\begin{subfigure}{0.45\textwidth}
  \includegraphics[width=\linewidth]{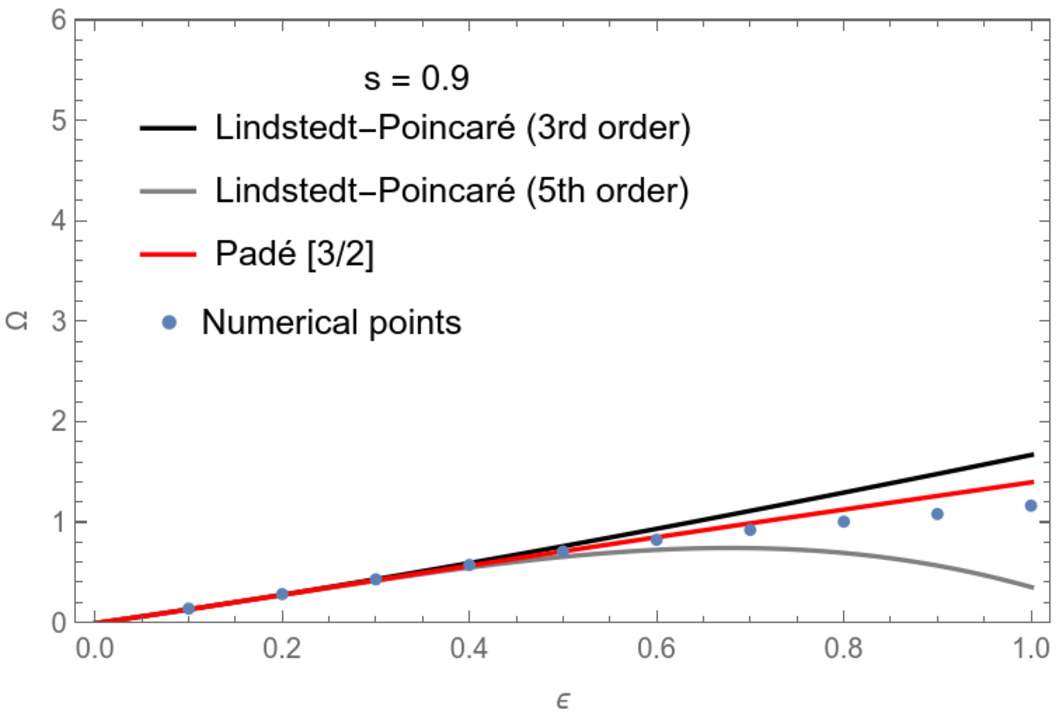}
  \caption{}
  \label{fig:0.9LPP}
\end{subfigure}\hfil 
\caption{Angle of deviation as a function of $\epsilon$ for different spin parameters. Comparison of equations (\ref{eq:lindstedtSol}), (\ref{eq:lindstedt5}), and different orders of Pad\'e approximants with the numerical solution to equation (\ref{eq:orbit2}). These plots consider the spin parameters and the mean statistical error for the Pad\'e approximation: 
(a) $s=-0.6$, $e=1.83\%$, (b) $s=-0.3$, $e=2.02\%$, (c) $s=0$, $e=3.29\%$, (d) $s=0.3$, $e=3.24\%$, (e) $s=0.6$, $e=5.25\%$, and (f) $s=0.9$, $e=6.48\%$.}
\label{fig:lindst-pade}
\end{figure}

From figures (\ref{fig:-0.9LPP}) and (\ref{fig:lindst-pade}) it can be observed that the error increases for higher spin values, also, when considering points near $\epsilon=1$. This is expected, since the perturbation method works for small values of $\epsilon$, and the Pad\'e approximant is derived from the Lindstedt-Poincar\'e solution, nonetheless, the Pad\'e approximant for each case gives a reasonable approximation. The mean statistical error for the Pad\'e approximants with respect to the numerical solutions is between $2\% - 7\%$ for different spin parameters and Pad\'e expressions. Therefore, we have found expressions that correctly describe the behavior of photons deviating their trajectory due to the gravitational field of a rotating black hole.

\FloatBarrier

\section{Equation of the Orbit for Arbitrary Angles}

\begin{figure}[htb]
    \centering
\begin{subfigure}{0.5\textwidth}
  \includegraphics[width=\linewidth]{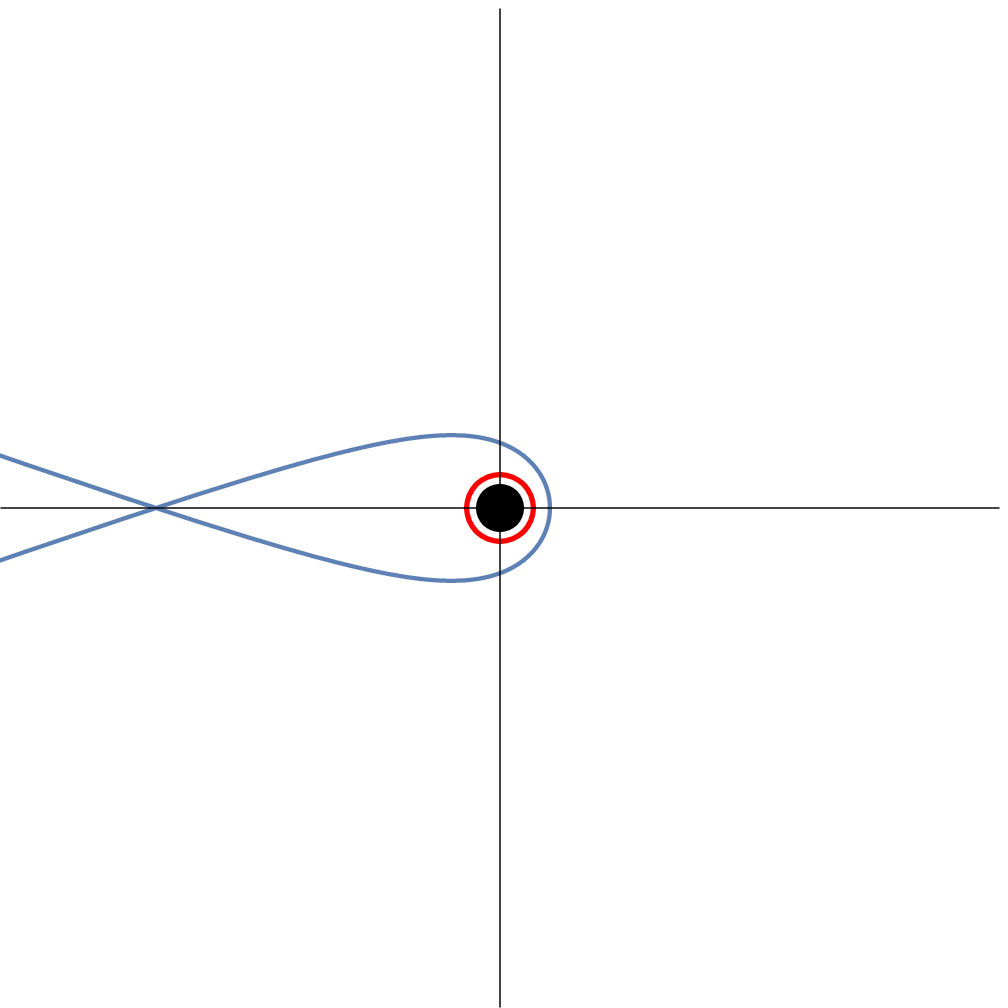}
  \caption{}
  \label{fig:retro1}
\end{subfigure}\hfil
\begin{subfigure}{0.5\textwidth}
  \includegraphics[width=\linewidth]{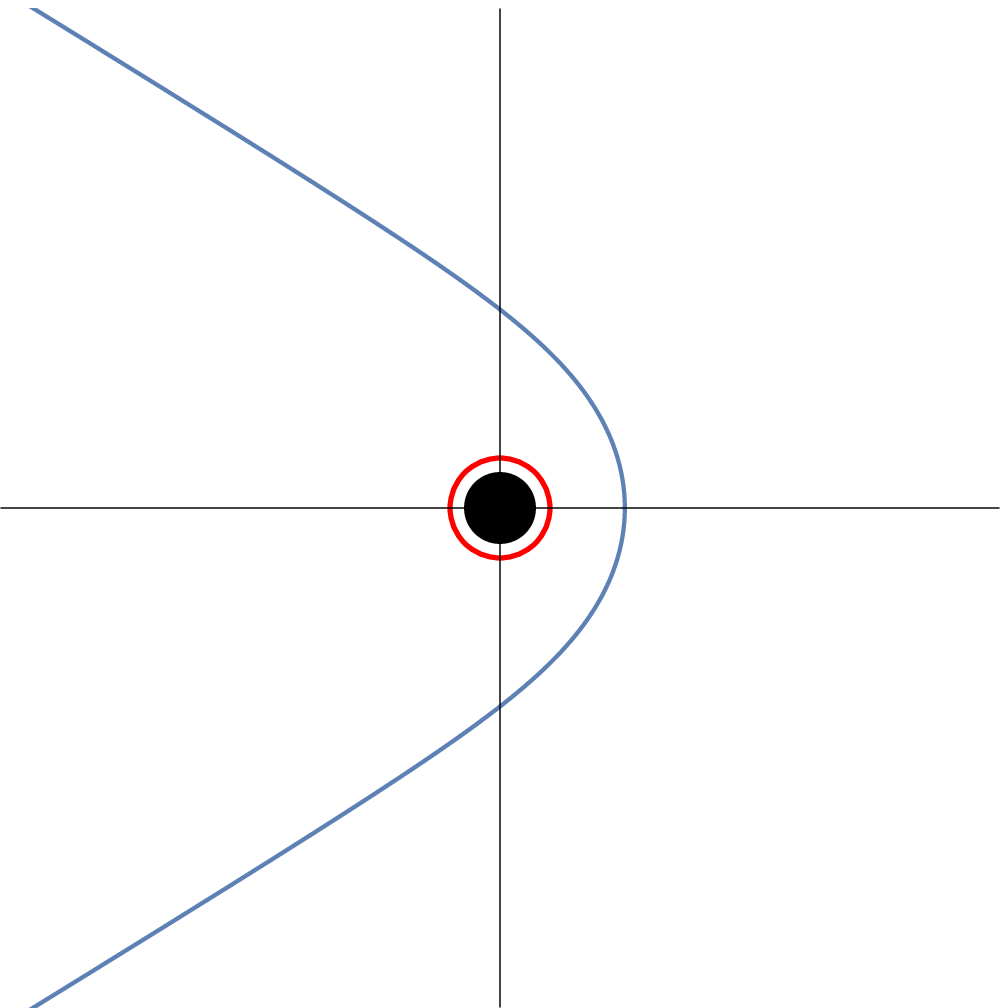}
  \caption{}
  \label{fig:retro6}
\end{subfigure}\hfil

\medskip
\begin{subfigure}{0.5\textwidth}
  \includegraphics[width=\linewidth]{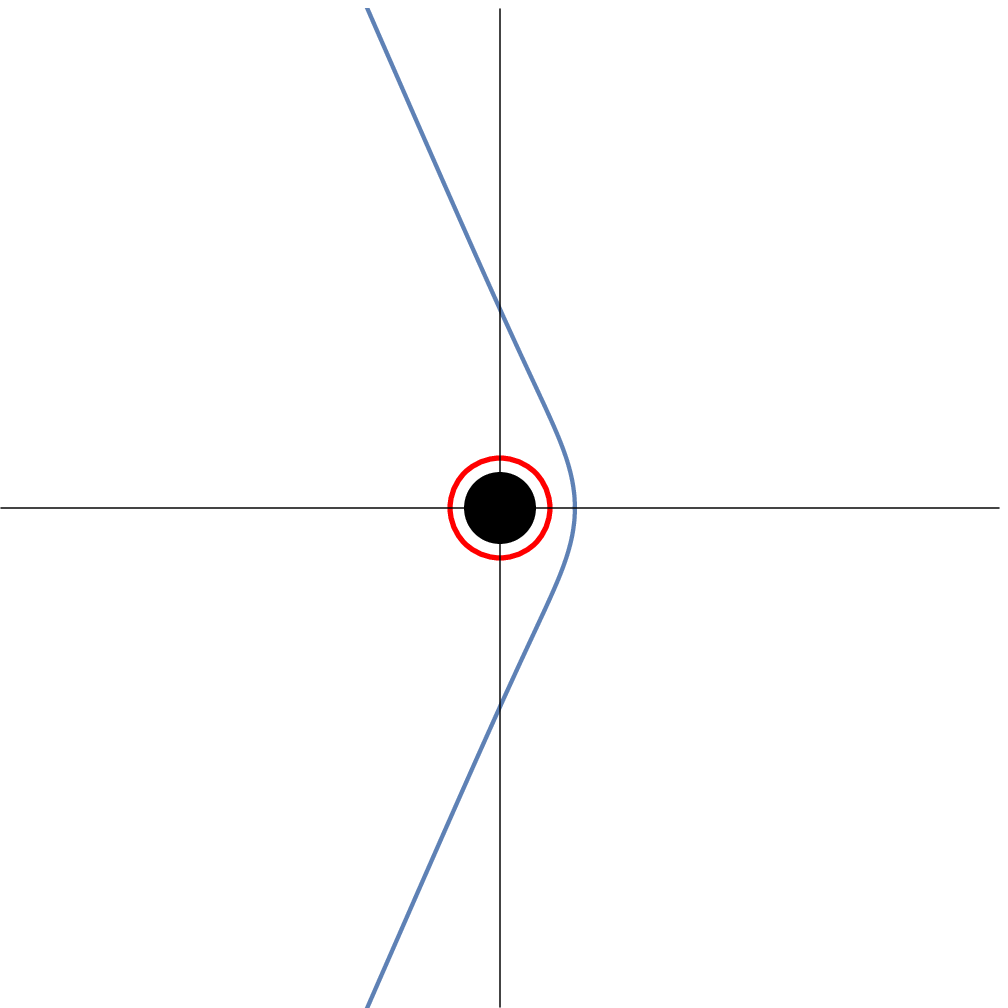}
  \caption{}
  \label{fig:direct1}
\end{subfigure}\hfil
\begin{subfigure}{0.5\textwidth}
  \includegraphics[width=\linewidth]{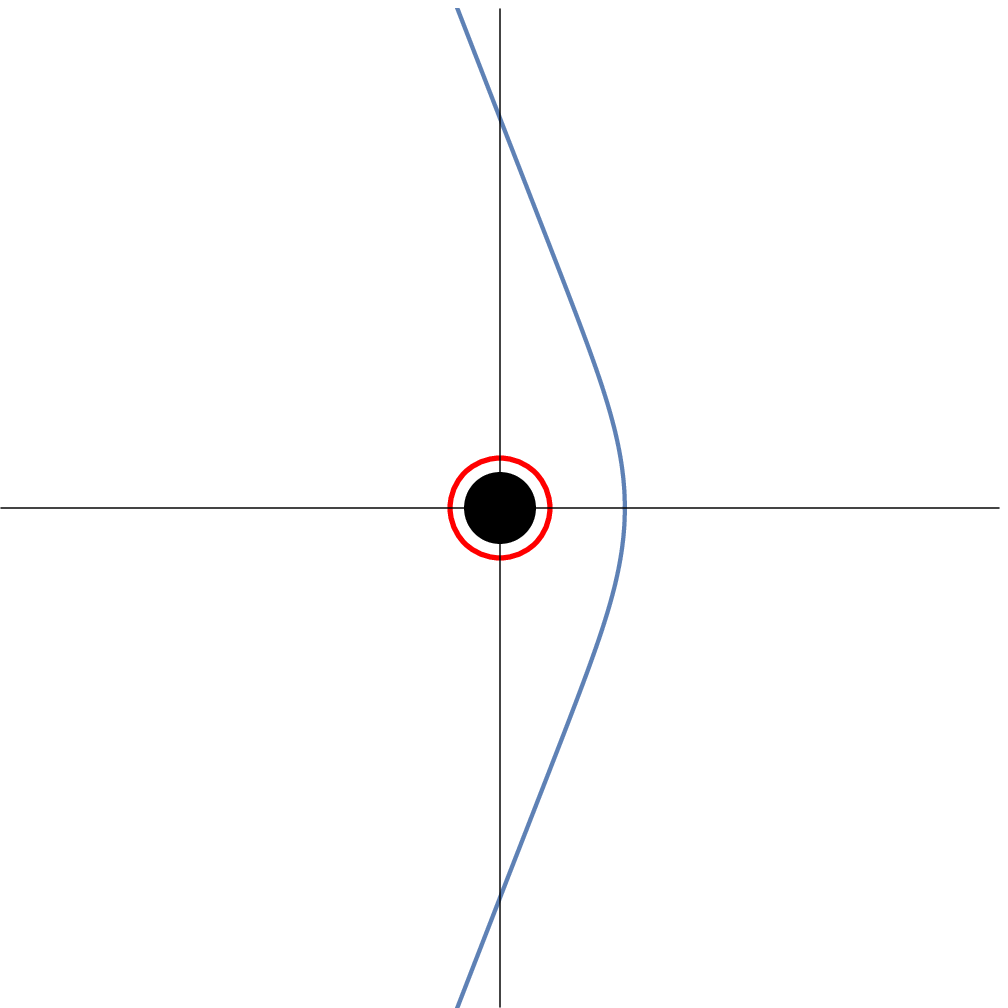}
  \caption{}
  \label{fig:direct6}
\end{subfigure}\hfil
\caption{Trajectories of photons in the equatorial plane of a Kerr black hole for different $\epsilon$. The black disk represents the outermost event horizon and the red circumference is the ergosphere. Each figure considers a:  
(a) retrograde orbit for $s=-0.9$ and $\epsilon=1$, (b) retrograde orbit for $s=-0.9$ and $\epsilon=0.6$, (c) direct orbit for $s=0.9$ and $\epsilon=1$, and (d) direct orbit for $s=0.9$ and $\epsilon=0.6$. }
\label{fig:trajectories}
\end{figure}

The procedure followed to find the Lindstedt-Poincar\'e solutions for the angle of deviation, provides us with a simplified and well behaved equation of the orbit. Even though several simplifications were considered to get equation \ref{eq:orbitGeneral}, it correctly describes the trajectories of massless particles around rotating black holes. Solving said equation numerically leads to interesting examples in the vicinity of the ergosphere. Figure \ref{fig:trajectories} depicts the nature of the numerical solution analyzed in the previous section, and illustrates the behavior of the equation that was used to calculate the angle of deviation.

\FloatBarrier

\section{Conclusions}
\label{Conclusions}

In this paper we present a way to solve the equation of motion for null geodesics in the equatorial plane of the Kerr metric. Being particularly interested in one of the first and most exciting predictions of General Relativity, the deviation of light that passes in the vicinity of a strong gravitational field. We focus on rotating black holes, the total angular momentum carried by these enormous compact objects curves space-time in a very interesting way. The frame dragging effect that occurs around a Kerr black hole can indeed make the equations that describe the motion of particles, very complex; this is why we have simplified the metric to the equatorial plane.

As a first attempt to solve the equation of motion, we try a traditional perturbative treatment, with a small parameter $\epsilon=\frac{r_c}{b}$, but this method yields a problem for solutions of higher order than two. When trying to solve for third order, we find ourselves dealing with terms that grow boundlessly, this of course is unwanted behavior in our solution, convergence is necessary. These secular terms of the form $\phi \sin\phi$ are oscillating with a growing amplitude, which may lead to non-uniformity in the solutions; additionally, difficulty in solving equations of order $n$ arises. To go around this issue, we applied the Lindstedt-Poincar\'e method, which expands the variable that appears in the secular terms, this allows for adequate behavior once the coefficients are chosen correctly such that the secular term is eliminated. As it can be observed in the figure \ref{fig:lindst}, this method preserves the behavior of the numerical solution, yet it lacks precision. Finally, in an attempt to further increase precision in the approximation of the deviation angle, Pad\'e approximants were calculated from the result of the Lindstedt-Poincar\'e method. It was necessary to calculate a fifth order solution with the perturbative method (equation \ref{eq:lindstedt5}) in order to acquire good approximations for the angle of deviation. In figures \ref{fig:-0.9LPP} and \ref{fig:lindst-pade}, it can be observed that the Pad\'e expressions increase precision for the angle of deviation. In previous work \cite{Rodriguez-Marin, Marin-Poveda}, it was demonstrated that the Pad\'e approximants produce better results than Lindstedt-Poincar\'e in the Schwarzschild and Reissner-Nordstrom metrics. This study shows that this is also the case for the Kerr metric. From figures \ref{fig:-0.9LPP} and \ref{fig:lindst-pade} we can see that the Lindstedt-Poincar\'e method will produce better results for small $\epsilon$, which is expected from the perturbative nature of the solution. In order to approximate the angle for regions closer to the black hole, the Pad\'e method was employed, producing favorable results. It may be possible to find better approximations with higher order solutions. To calculate the higher order terms for the Lindstedt-Poincar\'e method, a similar procedure as the one shown in section \ref{LP} can be followed, the same goes for the Pad\'e approximants in section \ref{pade}.

\section*{Conflict of interest}
The authors declare no conflict of interest.


\appendix

\section{Higher order Pad\'e approximants}

In this appendix we will present the Pad\'e approximants that were used in section \ref{analysis}, calculated numerically from the fifth order Lindstedt-Poincar\'e solution (equation \ref{eq:lindstedt5}).

\medskip
\begin{equation}
    \Omega^{[2/1]}(\epsilon,s)=\frac{\frac{4 \epsilon }{3}+\frac{\left(-39296+3375 \pi ^2+3840 s+2400 \pi  s+3840 s^2\right) \epsilon ^2}{540 (-16+15 \pi -16 s)}}{1+\frac{2
(-1348+225 \pi -120 s+300 \pi  s) \epsilon }{45 (-16+15 \pi -16 s)}}
\end{equation}

\begin{equation}
    \begin{split}
        \Omega^{[1/4]}(\epsilon,s)&=4 \epsilon \left(3 \left(1+\frac{1}{48} (16-15 \pi +16 s) \epsilon \right.\right.\\
        &+\left.\left.\frac{\left(-39296+3375 \pi ^2+3840 s+2400 \pi  s+3840 s^2\right)\epsilon ^2}{34560}\right.\right.\\
        &+\left.\left. \frac{\epsilon ^3}{1658880}\left(1497088+643920 \pi -54000 \pi ^2-50625 \pi ^3+1746944 s \right.\right.\right.\\
        &+\left.\left.\left.76800 \pi  s-126000 \pi ^2 s-61440 s^2+134400 \pi  s^2+61440s^3\right)\right.\right.\\
        &+\left.\left.\frac{\epsilon ^4}{8360755200}\left(-3765628928-2318400000 \pi -1009260000 \pi ^2\right.\right.\right.\\
        &+\left.\left.\left.170100000 \pi ^3+79734375 \pi ^4-8797716480s-3672614400 \pi  s\right.\right.\right.\\
        &-\left.\left.\left.30240000 \pi ^2 s+340200000 \pi ^3 s+4035870720 s^2+645120000 \pi  s^2\right.\right.\right.\\
        &-\left.\left.\left.262080000 \pi ^2 s^2-309657600 s^3+387072000 \pi  s^3+103219200s^4\right) \right)\right)^{-1}
    \end{split}
\end{equation}

\newpage
\begin{equation}
    \begin{split}
        \Omega^{[2/3]}(\epsilon,s)&=\left(\frac{4 \epsilon }{3}+\left(\left(-3765628928-2318400000 \pi -1009260000 \pi^2\right.\right.\right.\\
        &+\left.\left.\left.170100000 \pi ^3+79734375 \pi ^4-8797716480 s-3672614400\pi  s\right.\right.\right.\\
        &-\left.\left.\left.30240000 \pi ^2 s+340200000 \pi ^3 s+4035870720 s^2+645120000 \pi  s^2\right.\right.\right.\\
        &-\left.\left.\left.262080000 \pi ^2 s^2-309657600 s^3+387072000 \pi  s^3+103219200 s^4\right)\epsilon ^2\right)/\right. \\
        &\left.\left(3780 \left(-1497088-643920 \pi +54000 \pi ^2+50625 \pi ^3-1746944 s\right.\right.\right.\\
        &\left.\left.\left.-76800 \pi  s+126000 \pi ^2 s+61440 s^2-134400 \pi  s^2-61440s^3\right)\right)\right)\left(1+\right.\\
        &\left.\left(\left(-392546048-65142000 \pi +5977125 \pi ^2+10631250 \pi ^3-890480640 s\right.\right.\right.\\
        &-\left.\left.\left.133249200 \pi  s+24570000 \pi ^2s+14175000 \pi ^3 s+75264000 s^2\right.\right.\right.\\
        &+\left.\left.\left.12096000 \pi  s^2+10080000 \pi ^2 s^2-19353600 s^3+16128000 \pi  s^3\right)\epsilon\right)/\right.\\
        &\left.\left(315 \left(-1497088-643920\pi +54000 \pi ^2+50625 \pi ^3-1746944 s\right.\right.\right.\\
        &-\left.\left.\left.76800 \pi  s+126000 \pi ^2 s+61440 s^2-134400 \pi  s^2-61440s^3\right)\right)\right.\\
        &+\left.\left(\left(21972307968+12282149760\pi -1974672000 \pi ^2-704851875 \pi ^3\right.\right.\right.\\
        &+\left.\left.\left.14955008000 s+923462400 \pi  s-2927484000 \pi ^2 s+567000000 \pi ^3 s\right.\right.\right.\\
        &-\left.\left.\left.11268096000 s^2-7545619200 \pi  s^2-584640000\pi ^2 s^2+604800000 \pi ^3 s^2\right.\right.\right.\\
        &+\left.\left.\left.1950842880 s^3+1032192000 \pi  s^3-645120000 \pi ^2 s^3-206438400 s^4\right) \epsilon^2\right)/\right.\\
        &\left.\left(15120 \left(-1497088-643920\pi +54000 \pi ^2+50625 \pi ^3-1746944 s\right.\right.\right.\\
        &-\left.\left.\left.76800 \pi  s+126000 \pi ^2 s+61440 s^2-134400 \pi  s^2-61440s^3\right)\right)\right.\\
        &+\left.\left(\left(-341247875072-434910873600\pi +1528905375 \pi ^2+34020000000 \pi ^3\right.\right.\right.\\
        &-\left.\left.\left.851421941760 s-523412121600 \pi  s+43337700000 \pi ^2 s+34955550000 \pi ^3 s\right.\right.\right.\\
        &-\left.\left.\left.1984223416320 s^2-514027584000\pi  s^2+147843360000 \pi ^2 s^2+27216000000 \pi ^3 s^2\right.\right.\right.\\
        &-\left.\left.\left.11302502400 s^3-288175104000 \pi  s^3-14515200000 \pi ^2 s^3+24192000000 \pi^3 s^3\right.\right.\right.\\
        &-\left.\left.\left.49545216000s^4+15482880000 \pi  s^4\right) \epsilon ^3\right)/\left(680400 \left(-1497088-643920 \pi \right.\right.\right.\\
        &+\left.\left.\left.54000 \pi ^2+50625 \pi ^3-1746944 s-76800 \pi  s+126000\pi ^2 s+61440 s^2\right.\right.\right.\\
        &-\left.\left.\left.134400 \pi  s^2-61440 s^3\right)\right)\right)^{-1}
    \end{split}
\end{equation}

\newpage
\begin{equation}
    \begin{split}
        \Omega^{[3/2]}(\epsilon,s)&=\left(\frac{4 \epsilon }{3}+\left(\left(21972307968+12282149760 \pi -1974672000 \pi ^2-704851875 \pi ^3\right.\right.\right.\\
        &+\left.\left.\left.14955008000 s+923462400 \pi  s-2927484000\pi ^2 s+567000000 \pi ^3 s-11268096000 s^2\right.\right.\right.\\
        &-\left.\left.\left.7545619200 \pi  s^2-584640000 \pi ^2 s^2+604800000 \pi ^3 s^2+1950842880 s^3\right.\right.\right.\\
        &+\left.\left.\left.1032192000 \pi  s^3-645120000\pi ^2 s^3-206438400 s^4\right) \epsilon ^2\right)/\left(252 \left(74054656\right.\right.\right.\\
        &+\left.\left.\left.11394000 \pi -6712875 \pi ^2-67522560 s+1966800 \pi  s+5400000 \pi ^2s-43223040 s^2\right.\right.\right.\\
        &-\left.\left.\left.2880000 \pi  s^2+5760000 \pi ^2 s^2+1843200s^3\right)\right)+\left(\left(341247875072+434910873600 \pi \right.\right.\right.\\
        &-\left.\left.\left.1528905375 \pi ^2-34020000000\pi ^3+851421941760 s+523412121600 \pi  s\right.\right.\right.\\
        &-\left.\left.\left.43337700000 \pi ^2 s-34955550000 \pi ^3 s+1984223416320 s^2+514027584000 \pi  s^2\right.\right.\right.\\
        &-\left.\left.\left.147843360000 \pi ^2 s^2-27216000000\pi ^3 s^2+11302502400 s^3+288175104000 \pi  s^3\right.\right.\right.\\
        &+\left.\left.\left.14515200000 \pi ^2 s^3-24192000000 \pi ^3 s^3+49545216000 s^4-15482880000 \pi  s^4\right) \epsilon^3\right)/\right.\\
        &\left.\left(11340 \left(74054656+11394000 \pi -6712875 \pi ^2-67522560 s+1966800 \pi  s+5400000 \pi ^2 s\right.\right.\right.\\
        &-\left.\left.\left.43223040 s^2-2880000 \pi  s^2+5760000\pi ^2 s^2+1843200 s^3\right)\right)\right)\left(1+\left(2 \left(945825920\right.\right.\right.\\
        &+\left.\left.\left.180704340 \pi -122590125 \pi ^2+490206336 s+297179400 \pi  s-102532500\pi ^2 s\right.\right.\right.\\
        &-\left.\left.\left.739737600 s^2-97171200 \pi  s^2+30240000 \pi ^2 s^2-83865600 s^3+16128000 \pi  s^3\right) \epsilon \right)/\right.\\
        &\left.\left(21 \left(74054656+11394000\pi -6712875 \pi ^2-67522560 s+1966800 \pi  s+5400000 \pi ^2 s\right.\right.\right.\\
        &-\left.\left.\left.43223040 s^2-2880000 \pi  s^2+5760000 \pi ^2 s^2+1843200 s^3\right)\right)+\left(\left(-40154343424\right.\right.\right.\\
        &+\left.\left.\left.7618126080\pi +1392490575 \pi ^2+179372756992 s+38285172480 \pi  s\right.\right.\right.\\
        &-\left.\left.\left.21447216000 \pi ^2 s+186239434752 s^2+38280883200 \pi  s^2-17160192000 \pi ^2 s^2\right.\right.\right.\\
        &-\left.\left.\left.23079813120s^3+7741440000 \pi  s^3+412876800 s^4\right) \epsilon ^2\right)/\left(1008 \left(74054656\right.\right.\right.\\
        &+\left.\left.\left.11394000 \pi -6712875 \pi ^2-67522560 s+1966800 \pi  s+5400000\pi ^2 s-43223040 s^2\right.\right.\right.\\
        &-\left.\left.\left.2880000 \pi  s^2+5760000 \pi ^2 s^2+1843200 s^3\right)\right)\right)^{-1}
    \end{split}
\end{equation}

\newpage
\begin{equation}
    \begin{split}
        \Omega^{[4/1]}(\epsilon,s)&=\left(\frac{4 \epsilon }{3}+\left(\left(-75609344+3909360 \pi +4644675 \pi ^2-58935296 s+4332720 \pi  s\right.\right.\right.\\
        &+\left.\left.\left.1512000 \pi ^2 s+10278912 s^2-806400\pi  s^2-860160 s^3\right) \epsilon ^2\right)\left(252 \left(-176000\right.\right.\right.\\
        &+\left.\left.\left.44235 \pi -191616 s+14400 \pi  s+7680 s^2\right)\right)^{-1}+\left(\left(-945825920-180704340 \pi\right.\right.\right.\\
        &+\left.\left.\left.122590125 \pi ^2-490206336 s-297179400 \pi  s+102532500 \pi ^2 s+739737600 s^2\right.\right.\right.\\
        &+\left.\left.\left.97171200 \pi  s^2-30240000 \pi ^2 s^2+83865600 s^3-16128000 \pi s^3\right) \epsilon ^3\right)/\left(5670 \right.\right.\\
        &\left.\left.\left(-176000+44235 \pi -191616 s+14400 \pi  s+7680 s^2\right)\right)+\left(\left(-40154343424\right.\right.\right.\\
        &+\left.\left.\left.7618126080\pi +1392490575 \pi ^2+179372756992 s+38285172480 \pi  s\right.\right.\right.\\
        &-\left.\left.\left.21447216000 \pi ^2 s+186239434752 s^2+38280883200 \pi  s^2-17160192000 \pi ^2 s^2\right.\right.\right.\\
        &-\left.\left.\left.23079813120s^3+7741440000 \pi  s^3+412876800 s^4\right) \epsilon ^4\right)/\left(544320 \left(-176000\right.\right.\right.\\
        &+\left.\left.\left.44235 \pi -191616 s+14400 \pi  s+7680 s^2\right)\right)\right)\cdot\\
        &\left(1+\frac{4\left(-1489396+427245 \pi -1564192 s+484680 \pi  s-161280 s^2\right) \epsilon }{21 \left(-176000+44235 \pi -191616 s+14400 \pi  s+7680 s^2\right)}\right)^{-1}
    \end{split}
\end{equation}

\end{document}